\documentclass[10pt,aps,prb,twocolumn,superscriptaddress, nolongbibliography]{revtex4-2}

\usepackage[usenames,dvipsnames,svgnames]{xcolor}

\usepackage{lmodern}
\usepackage[T1]{fontenc}
\setcitestyle{super}

\usepackage[utf8]{inputenc}
\usepackage{graphicx}
\usepackage{xcolor}

\usepackage{hyperref}
\hypersetup{colorlinks,allcolors=blue}

\begin{document}

\title{Mode-locked laser in nanophotonic lithium niobate}

\author{Qiushi Guo$^{1,2,3\dagger}$, Ryoto Sekine$^1$, James A. Williams$^1$, Benjamin K. Gutierrez$^4$, Robert M. Gray$^1$, Luis Ledezma$^{1, 5}$, Luis Costa$^1$, Arkadev Roy$^1$, Selina Zhou$^1$, Mingchen Liu$^1$, Alireza Marandi$^{1\dagger}$\\
\textit{$^1$Department of Electrical Engineering, California Institute of Technology, Pasadena, CA, USA\\$^2$Photonics Initiative, Advanced Science Research Center, City University of New York, NY, USA\\ $^3$Physics Program, Graduate Center, City University of New York, New York, NY, USA\\ $^4$Department of Applied Physics, California Institute of Technology, Pasadena, CA, USA \\* $^5$Jet Propulsion Laboratory, Pasadena, CA, USA\\}
$^\dagger$Email: \href{mailto:qguo@gc.cuny.edu}{qguo@gc.cuny.edu}; \href{mailto:marandi@caltech.edu}{marandi@caltech.edu}
}

\date{\today}

\begin{abstract}
Mode-locked lasers (MLLs) have enabled ultrafast sciences and technologies by generating ultrashort pulses with peak powers substantially exceeding their average powers. Recently, tremendous efforts have been focused on realizing integrated MLLs not only to address the challenges associated with their size and power demand, but also to enable transforming the ultrafast technologies into nanophotonic chips, and ultimately to unlock their potential for a plethora of applications. However, till now the prospect of integrated MLLs 
driving ultrafast nanophotonic circuits has remained elusive because of their typically low peak powers, lack of controllability, and challenges with integration with appropriate nanophotonic platforms. Here, we overcome these limitations by demonstrating an electrically-pumped actively MLL in nanophotonic lithium niobate based on its hybrid integration with a III-V semiconductor optical amplifier. Our MLL generates $\sim$4.8 ps optical pulses around 1065 nm at a repetition rate of $\sim$10 GHz, with pulse energy exceeding 2.6 pJ and a high peak power beyond 0.5 W. We show that both the repetition rate and the carrier-envelope-offset of the resulting frequency comb can be flexibly controlled in a wide range using the RF driving frequency and the pump current, paving the way for fully-stabilized on-chip frequency combs in nanophotonics. Our work marks an important step toward fully-integrated nonlinear and ultrafast photonic systems in nanophotonic lithium niobate. 
\end{abstract}

\maketitle 

\begin{figure*}[ht]
\centering
\includegraphics[width=0.95\linewidth]{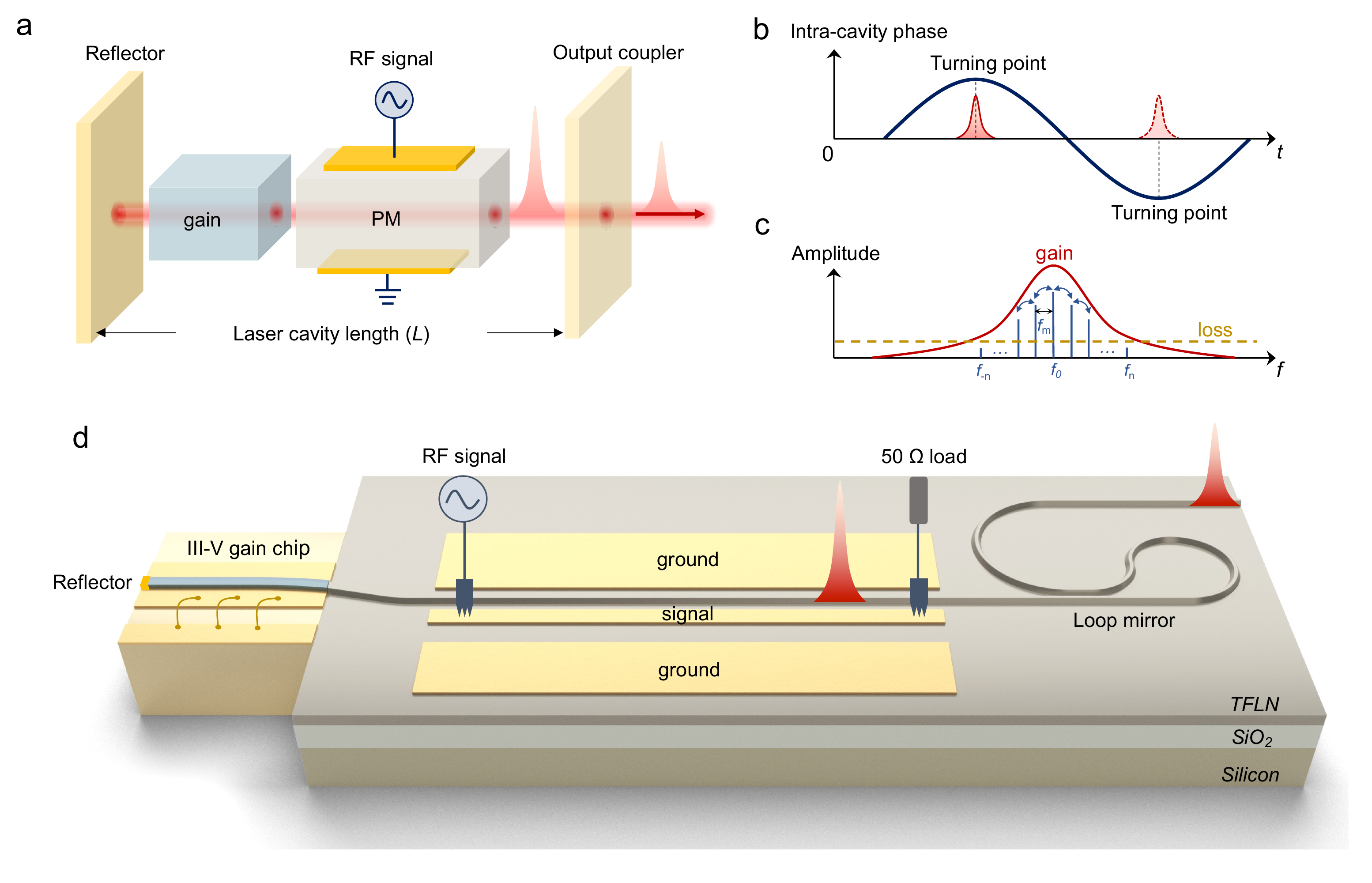}
\caption{\textbf{Principle and design of integrated actively MLL laser} \textbf{a,} Concept of active mode-locking through intra-cavity phase modulation. \textbf{b,} Illustration of mode-locking in the time domain. \textbf{c,} Illustration of mode-locking in the frequency domain. \textbf{d,} Schematic of the integrated actively MLL. The laser is composed of a single-angled facet (SAF) gain chip butt-coupled to a TFLN chip, which is composed of an electro-optic phase modulator and a broadband loop mirror. 
}\label{Fig1}
\end{figure*}

Mode-locked lasers (MLLs), which generate intense and coherent ultrashort optical pulses on picosecond and femtosecond timescales, have enabled numerous sciences and technologies in photonics such as extreme nonlinear optics\cite{wegener2005extreme}, femtochemistry\cite{zewail2000femtochemistry}, supercontinuum generation\cite{dudley2006supercontinuum}, optical atomic clocks \cite{diddams2001optical,ludlow2015optical}, optical frequency combs\cite{udem2002optical}, biological imaging\cite{piston1999imaging,horton2013vivo}, and photonic computing\cite{marandi2014network}. Today’s state-of-the-art MLLs are based on discrete fiber-based and free-space optical components and are expensive, power-demanding, and bulky. Realizing MLLs on integrated photonic platforms promises widespread utilization of ultrafast photonic systems which are currently limited to table-top laboratory experiments. However, so far the performance of integrated MLLs has not been on par with their table-top counterparts, lacking the required peak intensities and degrees of controllability required for on-chip ultrafast optical systems\cite{davenport2018integrated}, and many of the high-performance MLLs are not yet integratable with nanophotonic platforms. A major challenge lies in the simultaneous realization of large laser gain and an efficient mode-locking mechanism on integrated photonic platforms. Although III-V semiconductor gain media can be electrically pumped and they generally exhibit a very high gain per unit length and high saturation powers\cite{delfyett1992high}, the conventional method of achieving mode-locking and short pulse generation on the same semiconductor chip requires a narrow range of pumping current, thus significantly limiting the output power and the tunability of the integrated MLLs\cite{hermans2021high,vissers2022hybrid,bhardwaj2020monolithically}. 

To realize high-peak-power integrated MLLs, a promising approach consists of the hybrid integration of a semiconductor gain medium and an external mode-locking element based on electro-optic (EO) or nonlinear optical effects. Recently, thin-film lithium niobate (TFLN) has emerged as a promising integrated nonlinear photonic platform with access to power-efficient and high-speed EO modulation\cite{wang2018integrated,xu2022dual} and strong quadratic ($\chi^{(2)}$) optical nonlinearity\cite{wang2018ultrahigh,lu2019periodically}. Hybrid integration of semiconductor gain with TFLN enables a strong interplay between the laser gain and the EO or nonlinear effects to achieve active or passive mode-locking with high efficiency and tunability. Moreover, many of the nonlinear and ultrafast optical functionalities such as supercontinuum generation\cite{jankowski_ultrabroadband_2020}, optical parametric oscillation\cite{ledezma2022widely,lu2021ultralow,mckenna2022ultra,roy2022visible}, pulse shortening\cite{roy2022temporal}, all-optical switching\cite{guo2022femtojoule}, and quantum squeezing\cite{nehra2022few} can be realized in quasi-phase-matched LN nanophotonic devices with orders of magnitude lower peak powers compared to other platforms. Therefore, developing high-peak-power MLLs integrated into nanophotonic LN can enable a suite of nonlinear and ultrafast optical phenomena on a chip, promising integrated photonic systems with unprecedented performance and functionalities. 

In this work, we demonstrate a high-peak-power, electrically-pumped integrated actively MLL by hybrid integration of III-V semiconductors and LN nanophotonics. In contrast to conventional integrated MLLs based on hybrid integration of a III-V active region and a passive waveguide\cite{davenport2018integrated,hermans2021high}, our MLL synergistically exploits the high laser gain of III-V semiconductors and the efficient active optical phase modulation in LN nanophotonic waveguides as the mode-locking mechanism. Such a design eliminates the complexities associated with realizing gain and saturable absorption on the same semiconductor chip, allowing a much higher output power and a wider tunability of the laser. Under an external RF drive of less than 300 mW, our MLL generates ultrashort optical pulses around 1065 nm with a pulse duration of approximately 5 ps, pulse energy greater than 5 pJ, and a peak power greater than 0.5 W. This represents the highest reported peak power at a repetition rate of $\sim$10 GHz for integrated MLLs in nanophotonics. The MLL can operate over a broad range of electrical pumping currents and RF driving frequencies, and provide precise control of the carrier frequency and repetition rate of the resulting frequency comb, which can lead to fully-stabilized comb sources. In contrast to other recent demonstrations of ultrashort pulse sources on the TFLN platform such as Kerr soliton micro-combs\cite{gong2019soliton,he2019self} and electro-optic (EO) combs\cite{zhang2019broadband,hu2022high,yu2022integrated}, our MLL provides significantly higher on-chip peak power and pulse energies, and electrical-to-short-pulse efficiencies. Moreover, our MLL offers great system simplicity by eliminating the need for wavelength-tunable pump lasers, external optical amplification stages, complex cavity locking schemes, and/or pulse compression elements with high optical loss. The simplicity of our MLL design, combined with its high peak power, few-picosecond pulses, and controllable frequency comb parameters offers a practical path for fully-integrated nonlinear and ultrafast photonic systems in LN nanophotonics.

\begin{figure*}[ht]
\centering
\includegraphics[width=1\linewidth]{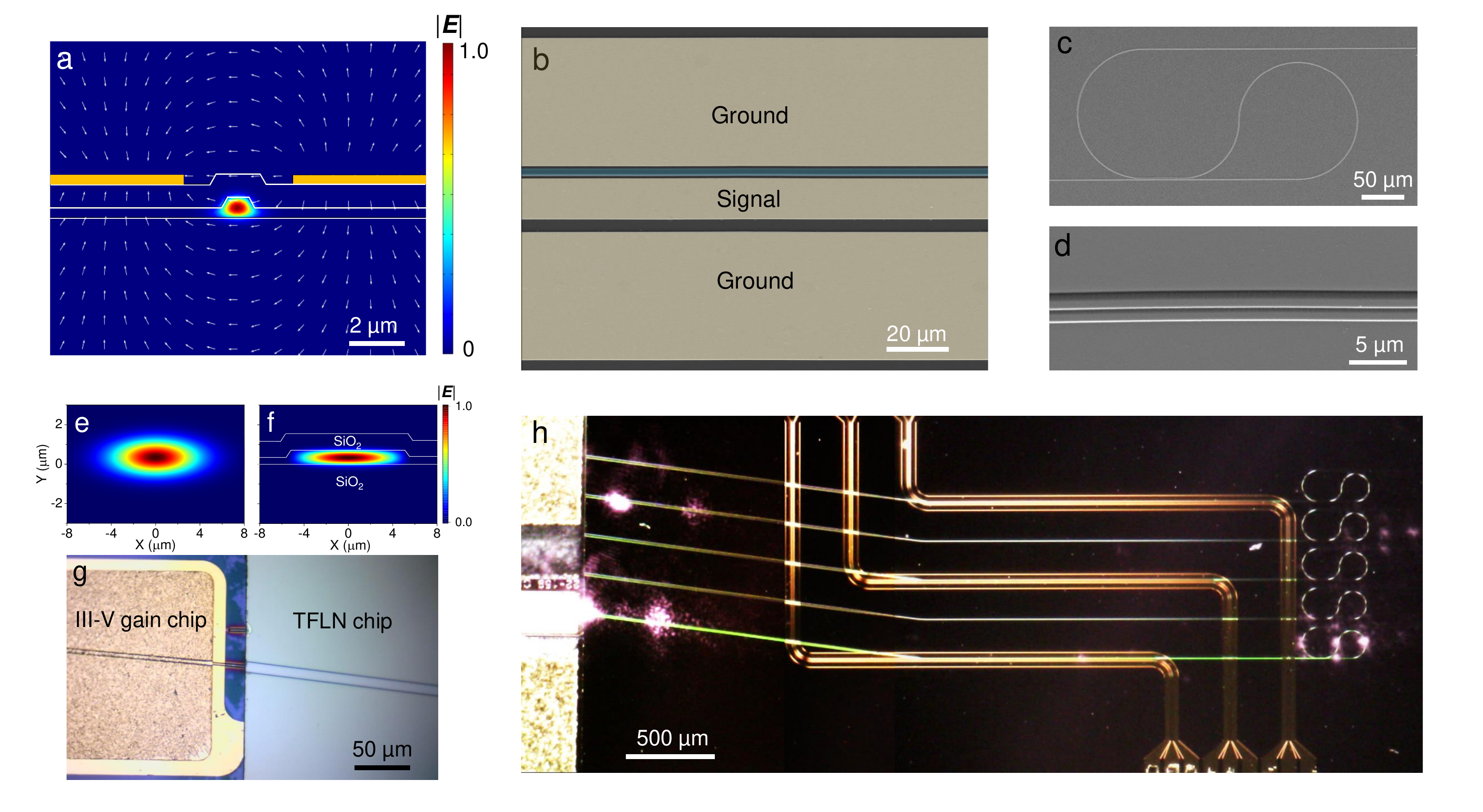}
\caption{\textbf{Integrated actively MLL laser on TFLN.} 
\textbf{a,} Cross-sectional view of the phase modulator (PM) region and the distribution of microwave field (white arrows) at 10 GHz and the optical field of the fundamental TE mode (color map) at 1065 nm. The top width of the TFLN waveguide is 800 nm. The gap between the signal and the ground electrodes is designed to be 4 $\mu$m. The RF electrodes are marked in yellow.  \textbf{b,} False-colored scanning electron microscope (SEM) image of the PM region in the fabricated device. The RF electrodes are marked in yellow and the optical waveguide is marked in blue. The remaining dark grey regions are the etched TFLN slab. \textbf{c,} SEM image of the broadband loop mirror. \textbf{d,} Zoom-in view of the curved coupling region of the broadband loop mirror. \textbf{e,} Fundamental TE mode profile at 1065 nm in the SAF gain chip waveguide. \textbf{f,} Fundamental TE mode profile at 1065 nm in the TFLN waveguide taper. \textbf{g,} Optical microscope image showing the coupling region between the two chips. \textbf{h,} Dark-field optical microscope image of the integrated actively MLL when operating.
}\label{Fig1}
\end{figure*}

\textbf{Operating principle and design of the MLL:} Figure 1a shows the concept of active mode-locking by electro-optic phase modulation inside a laser cavity. 
In the time domain, when a phase modulator (PM) is driven by a sinusoidal RF signal at a frequency $f_\mathrm{m}$, the intra-cavity phase modulation is equivalent to the cavity length modulation. Therefore, the laser cavity can be considered as having a moving end mirror with a sinusoidal motion at frequency $f_\mathrm{m}$. When an optical signal inside the cavity strikes this moving end mirror and gets reflected back, its optical frequency acquires a Doppler shift. After successive round trips, these Doppler shifts will accumulate, resulting in no steady-state solution. However, when a short circulating pulse strikes the end mirror at either of the ``turning points'' where the mirror reverses its direction (the extremum of the phase variation as shown in Fig. 1b), it will not acquire a Doppler frequency shift but instead a small quadratic phase modulation or chirp\cite{kuizenga1970fm,yu2022integrated}. Thus, an optical pulse can be maintained in the laser cavity after successive round trips as a steady-state solution. The characteristics of the pulse depend on the gain, loss, and dispersion in the cavity as well as the chirp from the PM. While in principle optical pulses can occur at either of the two phase modulation extrema and acquire chirp of different signs, the dispersion in the cavity can compensate for the chirp imposed by the PM at one extremum, and further chirp the pulse formed at another extremum. As a result, the combination of dispersion, gain, and nonlinear effect in the cavity can favor one pulse over the other, leading to only one pulse in the cavity\cite{nagar1992pure}. Such a mode-locking condition necessitates a good match between the phase modulation time period and the cavity round-trip time ($f_\mathrm{m}$ should be close to the cavity free spectral range (FSR)). The mode-locking mechanism can also be understood in the frequency domain. As shown in Fig. 1c, if the intra-cavity phase modulation frequency $f_\mathrm{m}$ matches the cavity FSR, the sidebands produced by each of the running axial modes are injected into the adjacent axial modes, resulting in the phase locking of adjacent modes. While the mutual injection of spectral modes within the cavity bears similarity to EO comb sources, it is important to note that in MLLs, these modes will lase due to the presence of laser gain within the cavity, whereas in EO comb sources, they are generated by dispersing the energy from a single pump laser line\cite{zhang2019broadband,yu2022integrated,hu2022high,rueda2019resonant}. This distinction gives rise to stark differences in their operations.


\begin{figure*}[ht]
\centering
\includegraphics[width=0.9\linewidth]{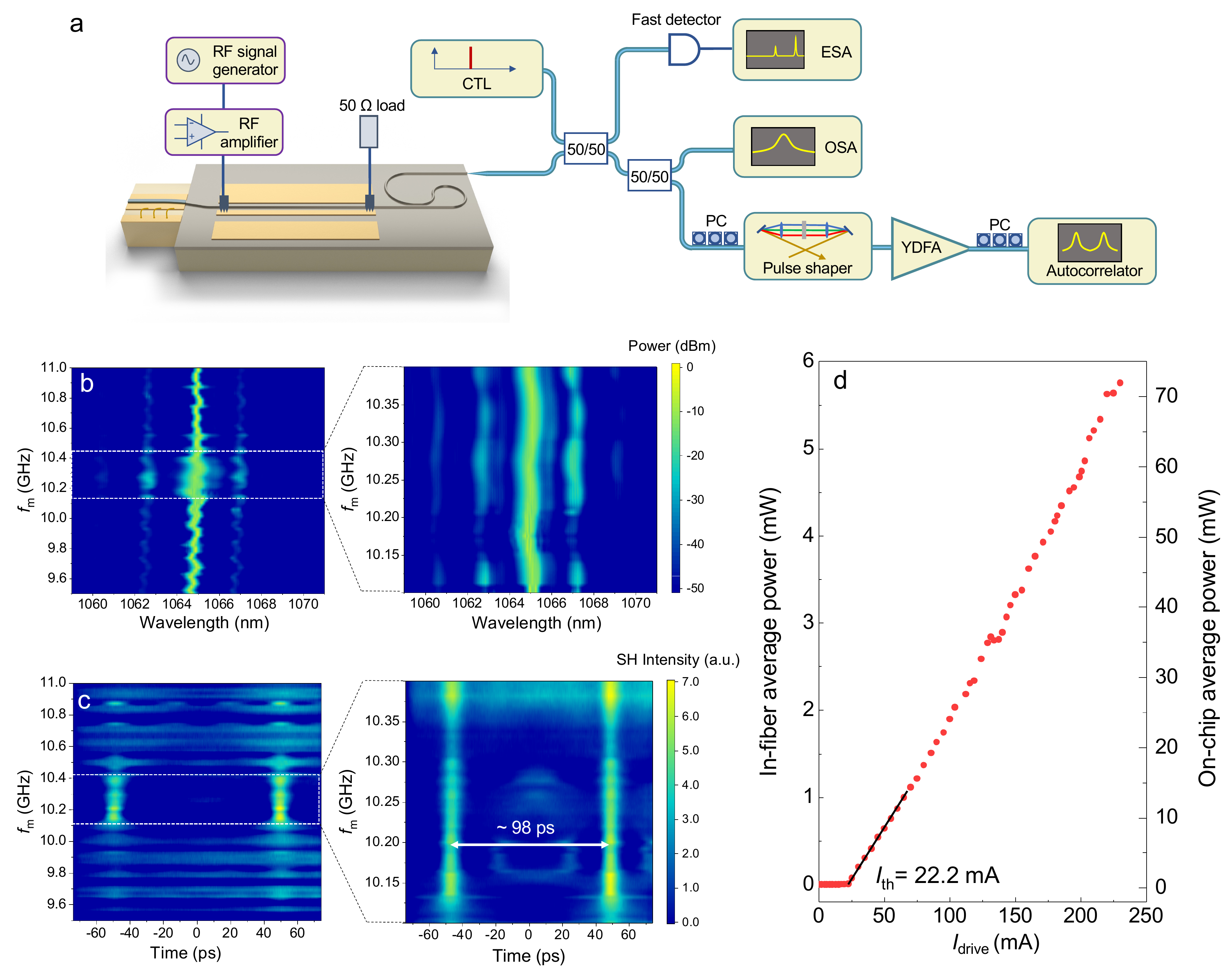}
\caption{\textbf{Characterization of integrated actively MLL and its operating regimes} \textbf{a}, Schematic of the setup for characterizing integrated actively MLL. The laser output from the waveguide facet was collected by a lensed fiber. In the measurement, we simultaneously monitored the optical spectrum from the optical spectrum analyzer (OSA), the intensity autocorrelation, and the heterodyne beatnote of the laser output. Heterodyne beat note was measured using a reference continuously tunable CW laser (Toptica CTL) operating around 1065 nm, a fast photodetector (FPD), and an electrical spectrum analyzer (ESA). \textbf{b}, The optical spectrum of the MLL output as a function of the RF driving frequency ($f_\mathrm{m}$). Significant spectrum broadening is found when $f_\mathrm{m}$ is between 10.1 and 10.4 GHz.  \textbf{c}, Intensity autocorrelation of the MLL output as a function of the $f_\mathrm{m}$. Two distinct intensity autocorrelation peaks separated by $\sim98$ ps emerged when $f_\mathrm{m}$ is tuned to be between 10.1 and 10.4 GHz. \textbf{d,} Dependence of laser average output power on the driving current ($I_\mathrm{drive}$) when $f_\mathrm{m}=10.17$ GHz.
}\label{Fig4}
\end{figure*}

Based on this principle, we design the integrated actively MLL as shown in Figure 1c. In our MLL, an electrically pumped gain section based on a single-angled facet GaAs gain chip (SAF gain chip) is butt-coupled to a TFLN chip, which contains an integrated EO phase modulator (PM) and a broadband loop mirror. A Fabry-Perot laser cavity configuration is formed between the reflective facet on the left end of the SAF gain chip and the broadband loop mirror on the TFLN chip. Here, an integrated PM is preferred over a Mach Zehnder interferometer (MZI)-based intensity modulator (IM) because the PM offers a lower insertion loss and avoids effects from the DC bias drift of the MZI modulator\cite{salvestrini2011analysis}. Thus, our MLL principle is distinct from that of an actively MLL based on an IM, in which the mode-locking is enabled by loss modulation\cite{perego2020coherent}. It is worth noting that semiconductor gain medium typically has a short carrier relaxation time (gain recovery time ($T_\mathrm{G}$)) on the order of ns\cite{delfyett1992high,duling1995compact,perego2020coherent}. To ensure the mode-locking and the formation of ultrashort optical pulses, $T_\mathrm{G}$ has to exceed the cavity round-trip time ($T_\mathrm{RT}$) of pulses by a large amount ($T_\mathrm{G}\gg T_\mathrm{RT}$)\cite{perego2020coherent}. In our design, by controlling the length of the TFLN waveguide, we realized this condition by having a cavity FSR of $\sim 10$ GHz, which translates to a cavity round-trip time of $\sim 100$ ps.

We fabricated our devices on a 700-nm-thick X-cut magnesium-oxide (MgO) doped TFLN on a SiO$_2$/Silicon (4.7 $\mu$m/500 $\mu$m) substrate (NANOLN). The details about the device fabrication can be found in the Methods. As shown in Fig. 2a, in the PM region, the RF electrodes are fabricated on top of the SiO$_2$ cladding layer. Such a design allows us to achieve high modulation efficiency (simulated value of 1.1 V$\cdot$ cm) by having a small gap (4 $\mu$m here) between the ground and signal electrodes and a significant overlap between the RF field and the optical field in the waveguide\cite{xu2022dual,jin2021efficient}. It also ensures a low optical propagation loss by offering a high tolerance to misalignment between the electrodes and the optical waveguide\cite{xu2022dual}. We designed the geometry of the RF electrode to ensure a 50 $\Omega$ impedance around 10 GHz. Figures 2b, c, and d show the scanning electron microscope (SEM) images of the PM and the loop mirror regions of the fabricated device. In the loop mirror, we adopted a curved coupling region design\cite{mitarai2020design} which increases the reflective bandwidth. The detailed design of the broadband loop mirror is described in the Supplementary Information Section II. Based on the length (1.5 mm) and the refractive index of the SAF gain chip around 1065 nm, we estimate that a $\sim$3-mm-long TFLN waveguide can lead to a laser cavity FSR of $\sim$10 GHz.

Figure 2e shows the 1065-nm fundamental TE mode profile in the waveguide of the SAF gain chip, which is calculated from the divergence angle of its emission. To minimize the coupling loss between the SAF gain chip and the TFLN chip, the top width of the input facet of the TFLN waveguide is tapered out to be 10.3 $\mu$m. The 1065-nm fundamental TE mode profile in the tapered TFLN waveguide is shown in Fig. 2f. This design ensures a maximal overlap with the optical mode produced by the SAF gain chip and yields a minimal coupling loss of $\sim$1.5 dB. In Section I of the Supplementary Information, we discuss the dependence of coupling loss on the lateral misalignment and gap between the SAF gain chip and TFLN waveguides.  The coupling loss can be further reduced by employing a polymer-based mode-size converter\cite{vissers2021hybrid}. Fig. 2g shows the microscope image of the coupling region after the alignment, in which the gap between the two chips is minimized. When the SAF gain chip is electrically pumped with a driving current ($I_\mathrm{drive}$) of 160 mA, we observe green light (the second harmonic of the 1065 nm light) inside the laser cavity (Fig. 1h), which indicates a high intra-cavity power around 1065 nm and a good alignment between the two chips.

\textbf{Characterization of the MLL:} We characterized the integrated actively MLL using an optical setup shown in Fig. 3a. We applied a $\sim$ 280 mW sinusoidal RF signal to the left end of the traveling wave electrodes (TWE) of the PM by the RF probe. The right end of the TWE is terminated by another RF probe with a 50 $\Omega$ load resistor mounted on it.  To investigate the operating regimes of the MLL, we simultaneously collect the laser output spectra, the intensity autocorrelation of the laser output in the time domain, and the heterodyne beat notes between two neighboring laser emission lines and a narrow-linewidth ($\sim$10 kHz) reference CW tunable laser (CTL, Toptica). In order to get intensity autocorrelation with a good signal-to-noise ratio, we pre-amplified the laser output power by a Ytterbium-doped fiber amplifier (YDFA). We also used a pulse shaper (Waveshaper 1000A, II-VI) to compensate for the group velocity dispersion (GVD) imposed by the phase modulator, YDFA, and the single-mode fiber. In the measurement, the gain chip is electrically pumped with an $I_\mathrm{drive}$ of 185.2 mA. 

As shown in Fig. 3b, when we widely scan the $f_\mathrm{m}$, the laser output exhibits a clear spectral broadening when $f_\mathrm{m}$ is between 10.1 and 10.4 GHz (labeled by the white dashed box). Meanwhile, within this $f_\mathrm{m}$ range, two distinct intensity autocorrelation peaks separated by $\sim$ 98 ps emerge (Fig. 3c), which indicates that optical pulses are formed in this $f_\mathrm{m}$ regime. At a $f_\mathrm{m}$ of 10.17 GHz, we measured the laser output power from the output facet of the TFLN chip with a single-mode lensed fiber. As shown in Fig. 3 d, the laser exhibits a very low threshold $I_\mathrm{drive}$ of 22 mA. Given the measured coupling loss of $\sim$11 dB between the TFLN waveguide and the single-mode lensed fiber, the on-chip laser output average power is more than 50 mW when the $I_\mathrm{drive}$ is greater than 180 mA.

\begin{figure*}[ht]
\centering
\includegraphics[width=1.02\linewidth]{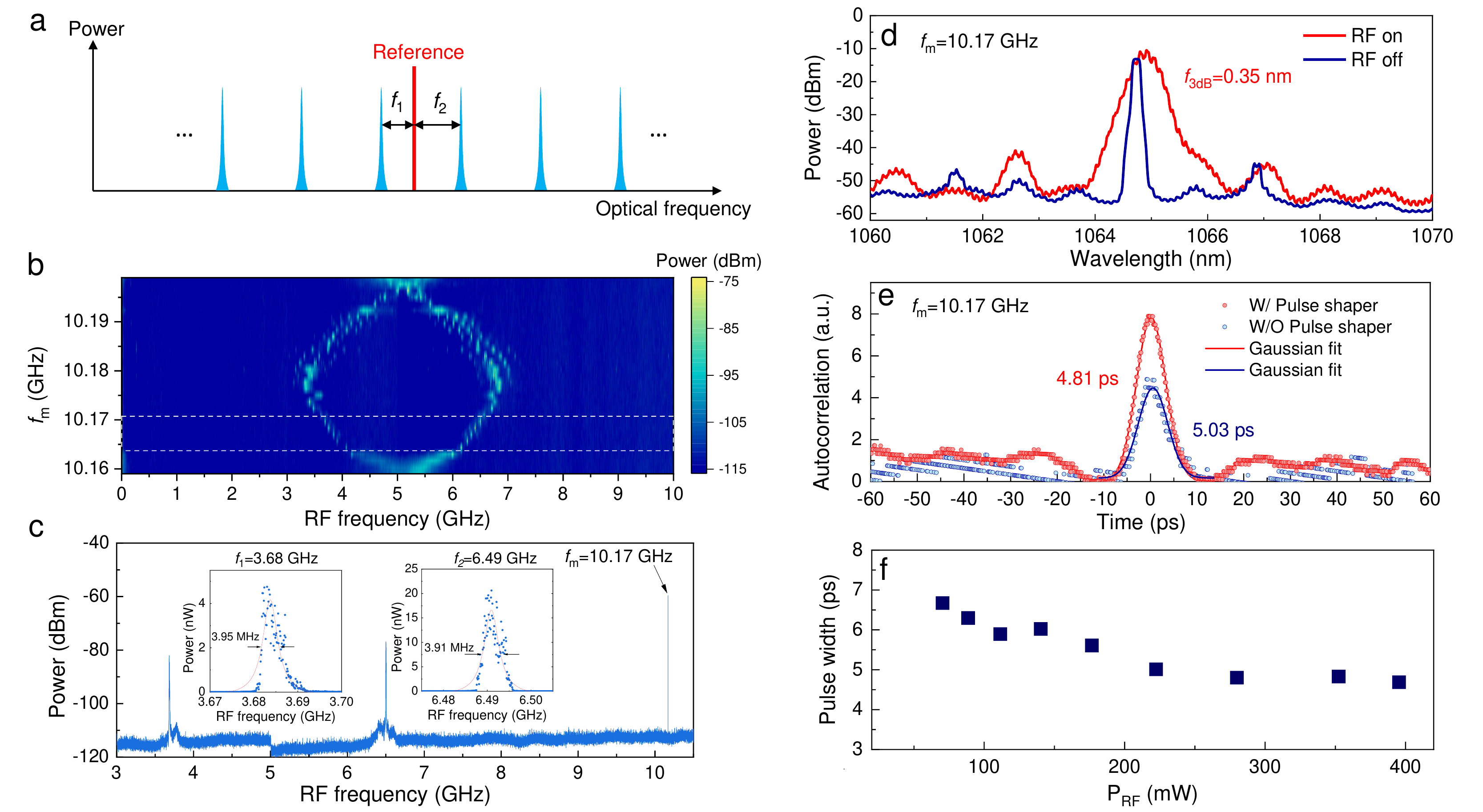}
\caption{\textbf{Finding the mode-locking regime of integrated actively MLL} 
\textbf{a}, Illustration of heterodyne beat notes generation. When the reference CTL frequency is between the two longitudinal mode frequencies of the MLL ouput near the center of its spectrum, two beat notes at RF frequencies of $f_1$ and $f_2$ will be detected by a fast detector \textbf{b}, Evolution of heterodyne beat notes with the $f_\mathrm{m}$. The mode-locking regime is marked by the white dashed box.  \textbf{c}, Heterodyne beat notes measured at $f_\mathrm{m}=10.17$ GHz. Insets: zoom-in view of the two beat notes at $f_1=3.68$ GHz and $f_2=6.49$ GHz. Blue symbols are measured data and solid red curves are Lorentz fits.  \textbf{d}, Output optical spectra of the MLL when the RF drive at 10.17 GHz is on (red) and off (blue). \textbf{e}, Intensity autocorrelation traces of the MLL output measured at $f_\mathrm{m}$ = 10.17 GHz with (red) and without (blue) the external pulse shaper. Symbols are measurement data and solid curves are Gaussian fits. \textbf{f}, Dependence of pulse width on the RF driving power ($P_\mathrm{RF}$).}\label{Fig3}
\end{figure*}

We further use the heterodyne beat notes to characterize the mode-locking and the resulting frequency comb. As illustrated in Fig. 4a, when the frequency of the reference CTL is resting in between the two neighboring comb lines of the MLL near the center of its spectrum, two RF beat notes at $f_1$ and $f_2$ are generated on the fast detector. Figure 4b shows the evolution of heterodyne beat notes as a function of $f_\mathrm{m}$.  When $f_\mathrm{m}$ is between 10.165 and 10.173 GHz as labeled by the white dashed box, two spectrally narrow beat notes are observed. This suggests that within this range of $f_\mathrm{m}$, the laser is operating in the mode-locked regime so that neighboring axial modes of the laser are phase-locked and it produces a frequency comb with narrow spectral peaks. In the time domain, this means a good synchronization between the round-trip of the pulse in the cavity and the phase modulation has been achieved, and the laser produces ultrashort optical pulses with high coherence. 

We also want to comment on some of the important behaviors of the resulting frequency comb when $f_\mathrm{m}$ is detuned from the cavity FSR. First, when the detuning is small (10.165 GHz<$f_\mathrm{m}$<10.173 GHz), we can still get two narrow beat notes, but $f_1$ and $f_2$ can shift significantly with $f_\mathrm{m}$, as shown in Fig. 4b. This indicates that the carrier frequency of the MLL sensitively depends on $f_\mathrm{m}$. Second, when the $f_\mathrm{m}$ is further detuned from the cavity FSR, the MLL exhibits a transition to a turbulent regime\cite{kartner1999turbulence}, which is manifested by multiple noisy beat notes around $f_1$ and $f_2$ in Fig. 4b. In the turbulent regime, the MLL can no longer reach steady state. In this regime, the laser can still emit ultrashort pulses as shown in Fig. 3c, albeit with low coherence.

As shown in Fig. 4c, at $f_\mathrm{m}$=10.17 GHz, we obtained two spectrally narrow RF beat notes at $f_\mathrm{1}$=3.68 GHz and $f_\mathrm{2}$=6.49 GHz, with a full width at half maximum (FWHM) linewidth of 3.95 MHz and 3.91 MHz, respectively. Given that the RF drive has a very small phase noise and no active locking of the laser cavity is used here, the linewidths of the heterodyne beat notes can be mainly limited by the drift of pulse carrier frequency. As shown in Fig. 4d, when a 280 mW RF drive at 10.17 GHz is applied to the PM, significant spectral broadening is observed. The pulse spectrum is centered at 1064.9 nm and the FWHM of the spectrum is 0.35 nm. Meanwhile, we also collected the intensity autocorrelation of the MLL output at $f_\mathrm{m}$=10.17 GHz, as shown in Fig. 4e. The autocorrelation trace indicates that the MLL produces one strong pulse at one of the modulation turning points, while the other pulse is significantly suppressed. The Gaussian fit of the intensity autocorrelation trace yields a pulse width of 4.81 ps (5.03 ps) with (without) the external pulse shaping. Since the pulse shaper can compensate for the chirp on the output pulse and the additional chirp imposed by the SMF and the YDFA, we expect the output pulse width directly after the MLL facet to be between 4.81 ps and 5.03 ps. The pulse width of 4.81 ps after pulse shaping corresponds to a time-bandwidth product of 0.445, which is very close to the transform-limited time-bandwidth product (0.44) of a Gaussian pulse\cite{siegman1986lasers}. To conservatively estimate the pulse energy and peak power, we use the measured output average power of 53 mW at $I_\mathrm{drive}=185.2$ mA and assume both pulses exist in the cavity. Hence, the output pulse energy of our MLL is at least 2.6 pJ and the pulse peak power is greater than 0.51 W.

We further studied the limits of the output pulse width of our MLL. First, we measured how the pulse width changes with the RF power ($P_\mathrm{RF})$ applied to the PM. As shown in Fig. 4f, the measured pulse only slightly decreases with increasing $P_\mathrm{RF}$, which is in good agreement with the power scaling law according to the Haus Master Equation (HME)\cite{haus2000mode}. We also found that further increasing the RF power will not shorten the pulse significantly. Instead, it can lead to laser instability due to index modulation of the TFLN waveguide caused by RF heating. By using the HME and ignoring the GVD and nonlinear effects in the laser cavity, we estimate that the pulse width limit of our actively MLL is $\sim$ 2.3 ps (see Supplementary Information Section IV for details). The experimentally measured pulse width is wider likely due to several factors, including cavity GVD, and the self-phase modulation and nonlinear chirp of the pulses imposed by the dynamical refractive index variation of the III-V gain medium associated with gain depletion and partial gain recovery\cite{delfyett1992high}.

\begin{figure}[ht]
\centering
\includegraphics[width=1.01\linewidth]{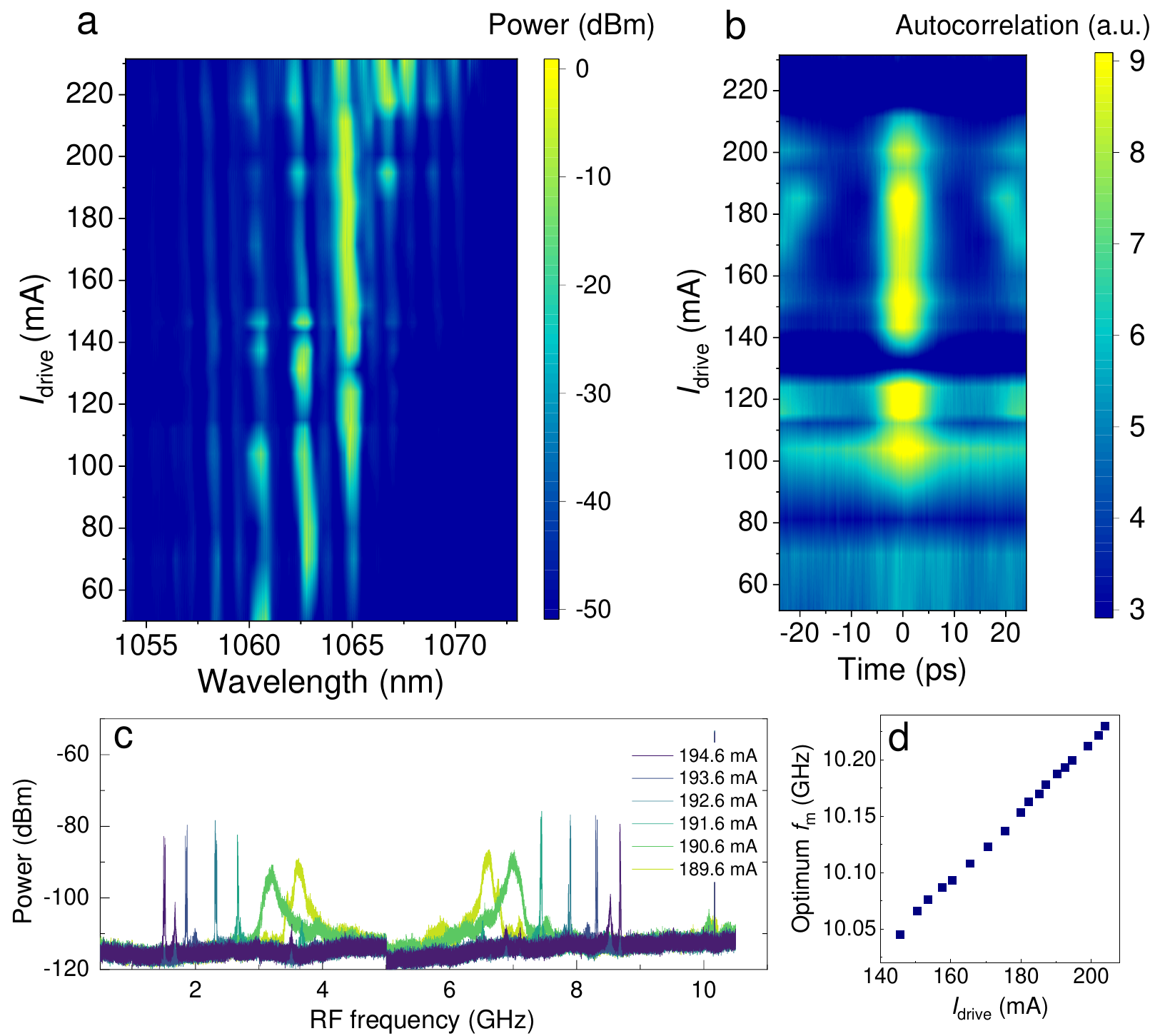}
\caption{\textbf{Curernt tuning of integrated actively MLL} 
\textbf{a}, The optical spectrum of the MLL output as a function of the driving current  ($I_\mathrm{mod}$). \textbf{b}, Autocorrelation trace of the MLL output as a function of $I_\mathrm{drive}$. In a and b, the 280 mW RF drive is fixed at 10.17 GHz.  \textbf{c}, Tuning of the heterodyne beat notes by the $I_\mathrm{drive}$. In this measurement, $f_\mathrm{m}$ is fixed at 10.18 GHz. \textbf{d}, Dependence of optimum $f_\mathrm{m}$ for mode-locking on $I_\mathrm{drive}$.
}\label{Fig5}
\end{figure}

\textbf{Current tuning of the MLL:} The electrical pumping current of the III-V gain chip ($I_\mathrm{drive}$) can serve as an important tuning knob of our MLL. Since $I_\mathrm{drive}$ can alter the gain spectrum and the refractive index of the gain medium, it can in turn vary the carrier frequency, the coherence property, and the repetition rate ($f_\mathrm{rep}$) of the MLL, and potentially lead to locking of the carrier frequency by applying active feedback on $I_\mathrm{drive}$. Figure 5 a and b show the dependence of the output spectra and autocorrelation of the MLL on the $I_\mathrm{drive}$ with 280 mW RF drive fixed at 10.17 GHz. It is evident that within a wide range of $I_\mathrm{drive}$ (140 - 205 mA), optical pulses can be formed inside the laser. In addition, it can be seen in Fig. 5a that the carrier frequency of the MLL blueshifts by $\sim$0.3 nm as the $I_\mathrm{drive}$ is increased from 140 mA to 200 mA. This blueshift, which has also been observed in other reports\cite{de2021iii}, is likely caused by the blueshift of the peak wavelength of the gain spectrum due to band filling and screening effects induced by carrier injection\cite{schmitt1985theory}. 

We then investigated the effect of $I_\mathrm{drive}$ on the coherence property and the $f_\mathrm{rep}$ of the laser. We kept the RF drive fixed at 280 mW and 10.18 GHz, and monitored the change in heterodyne beat notes $f_\mathrm{1}$ and $f_\mathrm{2}$ as we slightly varied $I_\mathrm{drive}$. The measurement results are summarized in Fig. 5c. As the $I_\mathrm{drive}$ is tuned from 189.6 mA to 194.6 mA, the laser transitions from the turbulent regime to the mode-locked regime, and then back to the turbulent regime. These results suggest that, with a frequency-stable reference CW laser and active feedback on $I_\mathrm{drive}$, it may be possible to lock the carrier frequency of the MLL and operate the device as a stable frequency comb, as the $f_\mathrm{rep}$ of the MLL has already been locked by the external RF oscillator. As shown in Fig. 5d, when we widely vary $I_\mathrm{drive}$ from 144 to 204 mA, the optimum $f_\mathrm{m}$ that enables mode-locking with high coherence can be varied from 10.04 GHz to 10.23 GHz, indicating the repetition rate of the laser can also be adjusted by $\sim$200 MHz by the $I_\mathrm{drive}$. Moreover, the optimum $f_\mathrm{m}$ increases almost linearly with $I_\mathrm{drive}$, which results from an increase of the cavity FSR caused by carrier injection in the gain medium.

\textbf{Conclusion and outlook:} In summary, we have demonstrated an integrated actively MLL in nanophotonic LN operating around 1065 nm, which can generate $\sim$5 ps ultrashort optical pulses. We estimate that the MLL produces an output pulse energy $\sim$2.6 pJ and a high peak power greater than 0.5 W, representing the highest pulse energy and peak power of any integrated MLLs in nanophotonic platforms. In contrast to conventional integrated MLLs that integrate both the gain and mode-locking elements on the same III-V chip, our MLL design decouples these elements, resulting in a significantly wider current tuning range and reconfigurability. This, in turn, allows for a wide tuning range of the laser $f_\mathrm{rep}$ of $\sim$200 MHz and precise control of the laser's coherence properties.

Looking forward, the current tuning capability of our MLL indicates that by using a reference and implementing active feedback to the $I_\mathrm{drive}$, we can achieve simultaneous locking of the carrier frequency and $f_\mathrm{rep}$ of the MLL. This allows the MLL to operate as a stable frequency comb with locked carrier frequency offset ($f_\mathrm{CEO}$) and $f_\mathrm{rep}$. In addition, comprehensive theoretical modeling of the laser dynamics and the identification of the single-pulse operating regime of the laser are of great importance for obtaining even higher peak powers and shorter pulses. We envision that the semiconductor gain and LN nanophotonic mode-locking elements can be fully integrated into the same chip and a better optical coupling between the two platforms can be achieved by adopting an advanced flip-chip bonding process\cite{shams2022electrically} or heterogeneous integration process\cite{de2021iii,zhang2023heterogeneous}. Furthermore, seamless integration of our high peak power MLL with other $\chi^{(2)}$ nonlinear optical functionalities provided by quasi-phase-matched TFLN nanophotonic devices offer exciting opportunities for the development of photonic systems that have yet to be realized in nanophotonics, such as fully integrated supercontinuum sources, self-referenced frequency combs, visible/ultraviolet femtosecond lasers, and atomic clocks.

\section*{Methods}
\noindent \textbf{Device fabrication.} We fabricated the integrated waveguides, phase modulators, and broadband loop mirrors on a 700-nm-thick X-cut MgO-doped LN thin-film on 4.7-$\mu$m-thick SiO$_2$ on top of a silicon substrate (NANOLN). We first patterned the waveguides using e-beam lithography by employing Hydrogen Silsesquioxane (HSQ) as the e-beam resist. The designed top width of the LN waveguide is 800 nm. The LN layer was etched by 350 nm using Ar$^+$ plasma. This etching process yields a waveguide sidewall angle of $\sim$60$^{\circ}$. Next, we deposited an 800 nm SiO$_2$ cladding layer using plasma-enhanced chemical vapor deposition (PECVD).  Another e-beam lithography step was used to pattern the RF metal electrodes on top of the cladding layer, in which PMMA was used as the e-beam resist. Then, we deposited Cr/Au (10 nm/300 nm) using e-beam evaporation. Metal electrodes are formed after metal lift-off in acetone. Finally, the waveguide facets were mechanically polished to enable good light coupling efficiencies.\\

\noindent \textbf{Optical measurements.} For butt coupling the SAF gain chip and the TFLN chip together, the SAF gain chip was placed on a 6-axis nano-positioning stage (Thorlabs) and the TFLN chip was clamped on a fixed sample stage. The two chips could be visually aligned by using a microscope from above. After visual alignment, the alignment was further optimized by maximizing the output power measured by a power meter, which is related to the intra-cavity optical power of the laser. The output power of the MLLs is probed by a single-mode lensed fiber connected to an optical power meter (Thorlabs). The RF drive is provided by an RF signal generator (Rohde $\&$ Schwarz SMA100B) and is subsequently amplified by a high-power RF amplifier (Mini-Circuits ZVE-3W-183+). The input RF power is calibrated by an RF power meter (Ladybug). For the results in Fig. 3-5, the laser output spectra were collected by an optical spectrum analyzer (OSA) covering 600-1700 nm (Yokogawa AQ6370B) with a 0.01 nm resolution bandwidth. The RF spectra were collected by an electronic spectrum analyzer (Rohde $\&$ Schwarz FSW) with a 100 Hz resolution bandwidth.\\

\noindent \textbf{Numerical simulations.} The optical and RF field distributions shown in Fig. 1a were simulated by COMSOL Multiphysics. We also used commercial software (Lumerical Inc.) to solve for the waveguide modes in order to design the waveguide taper and obtain the dispersion characteristics of the waveguide. In the simulation, the anisotropic index of the LN was modeled by the Sellmeier equations\cite{zelmon1997infrared}.


\section*{Data Availability}
The data that support the plots within this paper and other findings of this study
are available from the corresponding author upon reasonable request.
\section*{Code Availability}
The computer code used to perform the nonlinear simulations in this paper is available from the corresponding author upon reasonable request.
\section*{Acknowledgements}
The device nanofabrication was performed at the Kavli Nanoscience Institute (KNI) at Caltech. The authors thank Prof. K. Vahala for loaning equipment. Q.G. thanks Dr. M. Xu for the helpful discussions. The authors gratefully acknowledge support from ARO grant no. W911NF-23-1-0048, NSF grant no. 1846273 and 1918549, AFOSR award FA9550-20-1-0040, and NASA/JPL. The authors wish to thank NTT Research for their financial and technical support.
\section*{Authors Contributions}
Q.G. and A.M. conceived the project; Q.G. fabricated the devices with assistance from R.S.. Q.G performed the measurements, numerical simulation, and analyzed the data. R.S., J.W., B.G., R.M.G., L.L., L.C., and S.Z. assisted with the  measurements. B.G., A.R., and M. L. helped with the numerical simulation and data analysis. Q.G. wrote the manuscript with inputs from all authors. A.M. supervised the project.
\section*{Competing Interests}
Q.G. and A.M. are inventors on a patent application (US patent application no.
17/500,425) that covers the concept and implementation of the actively mode-locked laser here. The remaining authors declare no competing interests.

\newpage
\bibliography{references}

\end{document}


\title{Supplementary Information for ``Integrated lithium niobate actively mode-locked lasers with watt-level peak power'' }

\author{
Qiushi Guo$^1$, Ryoto Sekine$^1$, James A. Williams$^1$, Benjamin K. Gutierrez$^2$, Robert M. Gray$^1$, Luis Ledezma$^{1, 3}$, Luis Costa$^1$, Arkadev Roy$^1$, Selina Zhou$^1$, Mingchen Liu$^1$ and Alireza Marandi$^{1,\dagger}$\\
\textit{
$^1$ Department of Electrical Engineering, California Institute of Technology, Pasadena, California 91125, USA. \\
$^2$ Department of Applied Physics, California Institute of Technology, Pasadena, California 91125, USA\\
$^3$ Jet Propulsion Laboratory, California Institute of Technology, Pasadena, California 91109, USA.\\
}\\
$^\dagger$Email: \href{mailto:marandi@caltech.edu}{marandi@caltech.edu}
}





\maketitle


\renewcommand*\contentsname{Table of Contents}

\tableofcontents
\newpage

\section{Coupling between the gain chip and the thin-film lithium niobate chip}

\noindent To maximize the power coupled from the semiconductor gain chip into the photonic integrated chip, it is pertinent to ensure the mode overlap between the source output and the chip waveguide is maximized. Implementing an adiabitc taper is one method to achieve this. This style of taper improves the mode overlap in one dimension, which can significantly increase the total mode overlap. Alas, the mode mismatch in the remaining dimension can not be accommodated for by such a simple adiabatic taper. For this reason, it is important to carefully select a source such that the mode mismatch in this restrained dimension is maximized to begin with. To select a proper source, an estimate for the mode size at the output of the source is needed. With this, it is possible to determine if the source is a good choice for the designed waveguide mode of the photonic chip. A commonly listed specification for commercial sources is the far-field divergence angle of the full-width half maximum intensity profile. The following discusses how to estimate the output mode size for the source from this specification. As the gain chips considered for this design all exhibited a single spatial mode, single-mode Gaussian profiles are assumed in this analysis. This estimation is then used to inform the choice of gain chip implemented in the design. Finally, mode overlap simulations are performed using Lumerical MODE to optimize the mode overlap in the unrestricted dimension. This straightforward process enables efficient coupling between the gain chip and photonic circuit.

\subsection{Single mode Gaussian beam givergence}
\noindent For Gaussian beam diffraction, the beam radius at the 1/e point of the electric field is given by,
\begin{equation}
    w^2(z) = \omega_0^2 \left[ 1+\left(\frac{\lambda z}{\pi w_0^2 n}\right)^2\right]
    \label{beam waist}
\end{equation}

\noindent where $n$ is the refractive index of the medium the beam propagates through, $\lambda$ is the free-space wavelength, $z$ is the axial distance from the beam's waist, $w_0$ is the waist radius. In the far field ($z \gg \frac{\pi w_0^2 n}{\lambda}$), we can simplify Eq. \ref{beam waist} to 
\begin{equation}
    w(z) \approx \frac{\lambda z}{\pi n w_0}
\end{equation}

\noindent Additionally, the far-field diffraction half-apex angle at the 1/e point of the electric field, $\theta_{e}$, can be defined as,
\begin{equation}
    \tan(\theta_{e}) = \frac{w(z)}{z}
\end{equation}

\noindent Then, we can express the waist (radius) of the 1/e point of the electric field as
\begin{equation}
    w_0 \approx \frac{\lambda}{\pi n \tan(\theta_e)}
\end{equation}

\subsection{Electric Field and Intensity Relationship}

Specification sheets for commercially available semiconductor gain chips typically list the FWHM of the far-field intensity of the laser beam. However, up until this point, the equations listed have been based on the 1/e point of the electric field. So there is a need to connect the equations above to the specification sheets of these gain chips. At a specified propagation distance, the Gaussian profiles of the electric field and intensity are related by, 
\begin{equation}
    E(r) = e^{-r^2/2\sigma^2}
\end{equation}
\begin{equation}
    I(r) \propto |E(x)|^2 = e^{-r^2/\sigma^2}
\end{equation}
where the magnitudes of the profiles have been neglected. To determine the relationship between the half apex angle of the electric field at the 1/e point and the full apex angle of the FWHM intensity profile, a relationship between the waist radii at each of these points must be established. The FWHM of the intensity profile becomes,
\begin{equation}
        \frac{1}{2} = e^{-r_i^2/\sigma^2}
\end{equation}
\begin{equation}
        ln(0.5) = -r_i^2/\sigma^2
\end{equation}
\begin{equation}
        r_i^2 = -ln(0.5) \sigma^2
\end{equation}
The 1/e point of the electric field profile becomes,
\begin{equation}
        \frac{1}{e} = e^{-r_e^2/2\sigma^2}
\end{equation}
\begin{equation}
        1 = r_e^2/2\sigma^2
\end{equation}
\begin{equation}
        r_e^2 = 2 \sigma^2
\end{equation}
The radii at which each of these profiles achieves a specific value is different, thus $r$ in equations 5 and 6 have been replaced with $r_i$ and $r_e$ in equations 7-9 and 10-12 respectively. The ratio of the FWHM intensity and 1/e electric field radii is
\begin{equation}
    \frac{r_i}{r_e} = \sqrt{ -ln(0.5)/2}
\end{equation}
Now, both of these radii will occur at some shared propagation distance z away from the beam waist, given as
\begin{equation}
z = \frac{r}{\tan(\theta)}
\end{equation}
So we can find the relationship between the radii and their half-apex angles as,
\begin{equation}
\frac{r_i}{\tan(\theta_i)} = \frac{r_e}{\tan(\theta_e)}
\end{equation}
\begin{equation}
\frac{1}{\tan(\theta_e)} = \frac{r_i}{r_e}\frac{1}{\tan(\theta_i)} 
\end{equation}
Note that $\theta_e$ is the half-apex angle of the 1/e point of the electric field and $\theta_i$ is the half-apex angle of the FWHM point of the intensity.
Gathering equations 4, 13 and 16, the waist of the electric field can be expressed as
\begin{equation*}
\omega_0 \approx \frac{\lambda}{\pi n \tan(\theta_e)} = \frac{\lambda}{\pi n}\frac{r_i}{r_e}\frac{1}{\tan(\theta_i)} 
= \frac{\lambda}{\pi n}\frac{\sqrt{ -ln(0.5)/2}}{\tan(\theta_i)} 
\end{equation*}
So finally, taking $n=1$, we can convert directly from the spec sheet parameter of divergence angle at FWHM of the intensity to the waist (radius) of the electric field as
\begin{equation}
\omega_0 \approx \frac{\lambda}{\pi\tan(\theta_i)}\sqrt{\frac{-ln(0.5)}{2}}
\end{equation}

\noindent With this estimation, we can now quickly evaluate which gain chip options have a high mode overlap in the restrained dimension with our chosen waveguide structure. 

\subsection{An example using a Photodigm SOA}
\begin{figure*}[h!]
\centering
\includegraphics[width=0.6\linewidth]{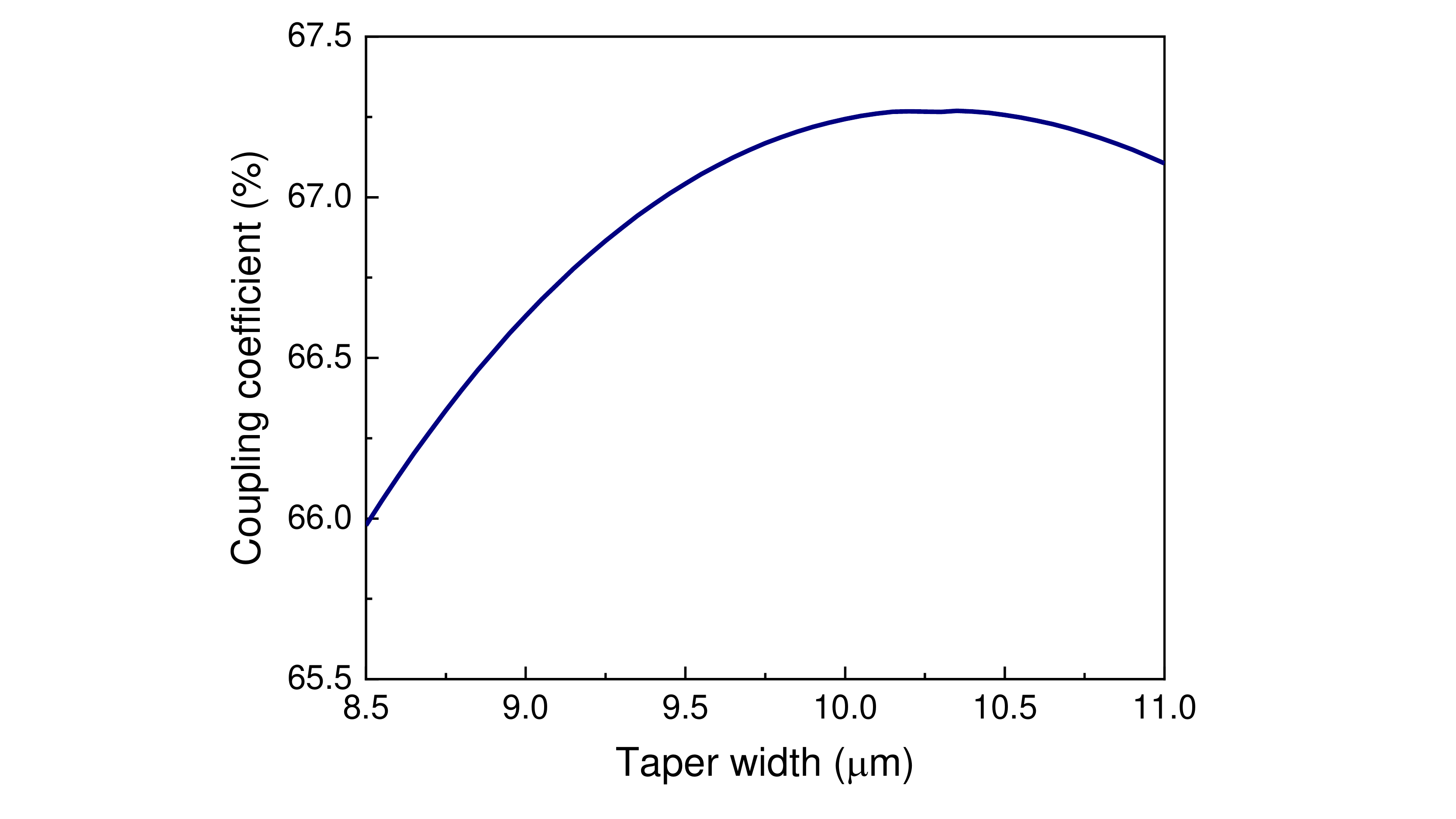}
\caption{Coupling efficiency as a function of the lateral offset between the gain chip and the TFLN chip.
}\label{fig:coupling ratio}
\end{figure*}

A Photodigm single-angle facet gain chip is listed as having a full-apex divergence angle at the full-width half-maximum (FWHM) point of the electric field as 26\textdegree \hspace{0.5mm} in the fast-axis (vertical) and  6\textdegree \hspace{0.5mm} in the slow-axis (horizontal). This gain chip emits around 1064 nm. Using $x$ to denote the horizontal axis and $y$ to denote the vertical axis, we can express the half-apex angles as $\theta_{i,x} = $3\textdegree \hspace{0.5mm} and $\theta_{i,y} = $13\textdegree. Using equation 17 we find the mode waist of the electric field at the 1/e point to be 
\begin{equation}
\omega_{0,x} \approx \frac{\lambda}{\pi\tan(\theta_{i,x})}\sqrt{\frac{-ln(0.5)}{2}} =
\frac{1064 \mathrm{nm}}{\pi\tan(3^{\circ})}\sqrt{\frac{-ln(0.5)}{2}} = 3.80\hspace{0.5mm} \mathrm{ \mu m}
\end{equation}
for the x axis and
\begin{equation}
\omega_{0,y} \approx \frac{\lambda}{\pi\tan(\theta_{i,y})}\sqrt{\frac{-ln(0.5)}{2}} =
\frac{1064 \mathrm{nm}}{\pi\tan(13^{\circ})}\sqrt{\frac{-ln(0.5)}{2}} = 863.63 \hspace{0.5mm} \mathrm{nm}
\end{equation} 
for the y-axis. The restricted dimension for this gain chip is found to have a high mode overlap with our chosen waveguide structure, making it a good choice for our application. An adiabatic taper can then be implemented to match the remaining horizontal dimension. To do this, mode overlap simulations are performed using Lumerical's finite-difference eigenmode (FDE) solver. The power coupling can then be calculated for various waveguide taper widths. To optimize the taper width, the power coupling is calculated as the taper width is swept, as shown in \ref{fig:coupling ratio}.

\begin{figure*}[h!]
\centering
\includegraphics[width=0.5\linewidth]{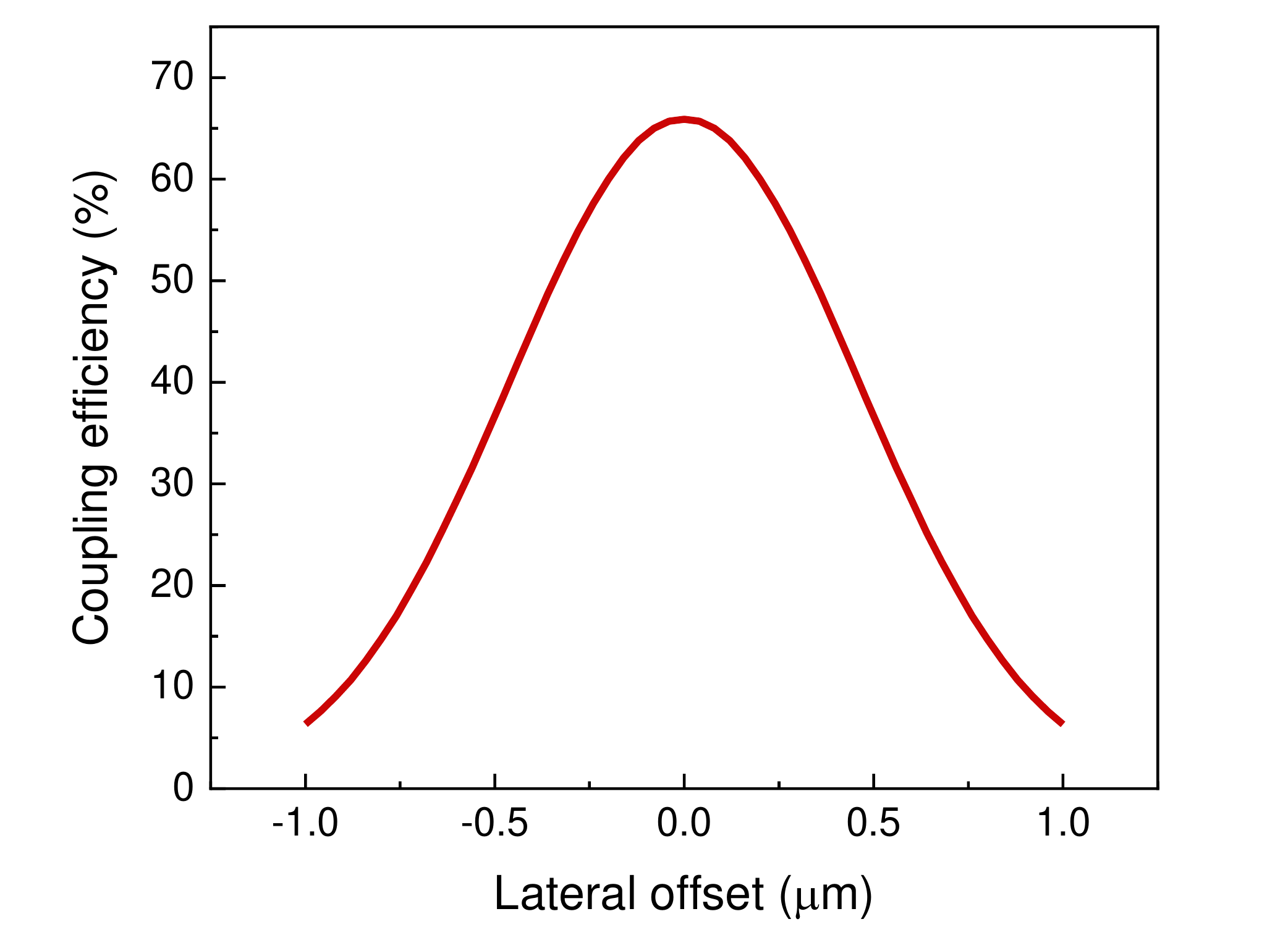}
\caption{Coupling efficiency as a function of the lateral offset between the gain chip and the TFLN chip.}\label{fig:coupler design}
\end{figure*}

For the Photodigm SOA, a taper width of 10.3 $\mu$m is found to have maximum power coupling. When the SOA is misaligned vertically from the adiabatic taper, the coupling efficiency drops off significantly, as can be seen in \ref{fig:coupler design}. Thus, while the adiabatic taper is a simple method to increasing coupling efficiency from the gain chip to our TFLN chip, its effectiveness is very sensitive to vertical alignment. 

\newpage
\section{Design of the broadband loop mirror}

\begin{figure*}[h!]
\centering
\includegraphics[width=0.7\linewidth]{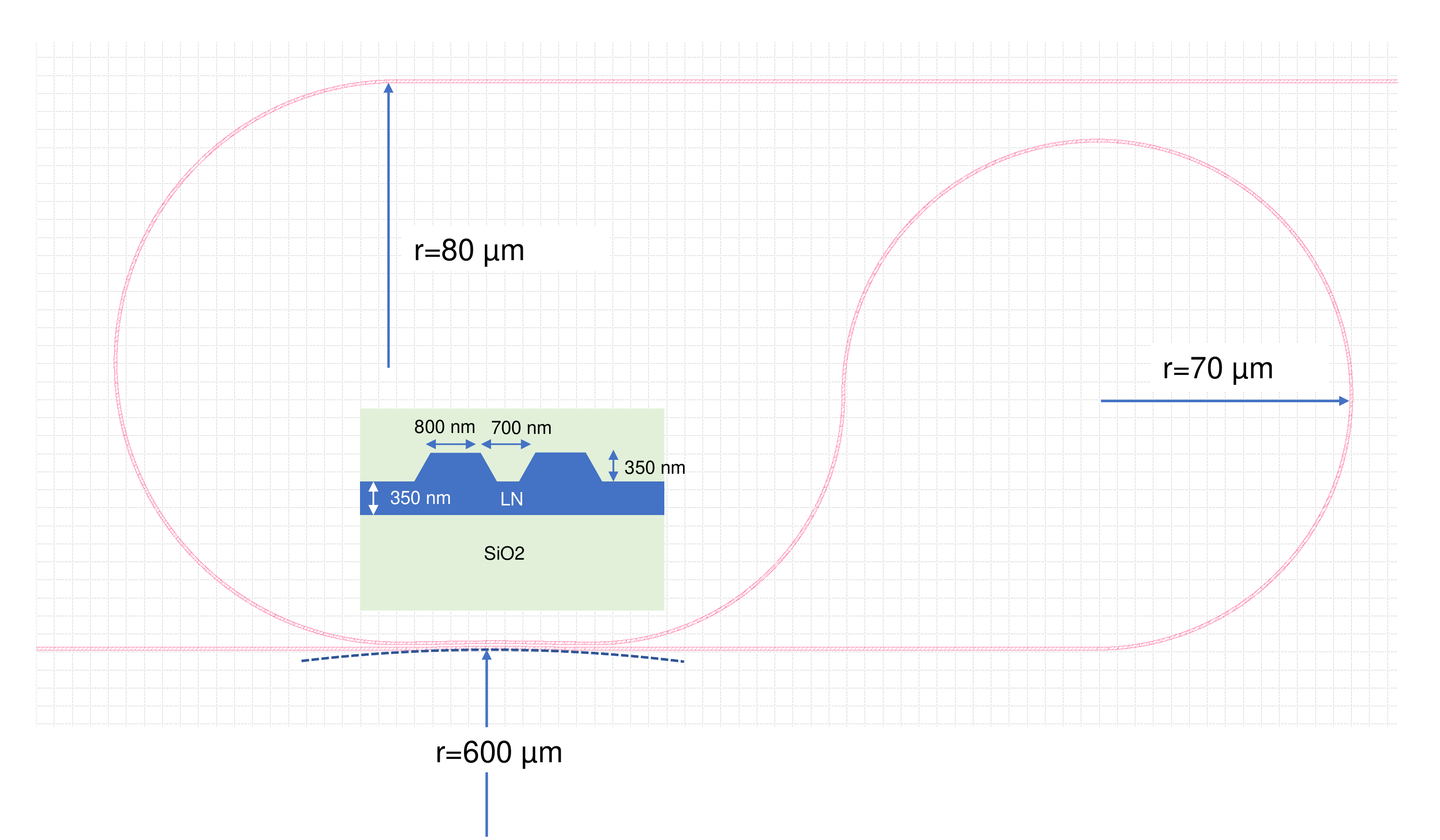}
\caption{Design of the broadband loop mirror with curved directional coupler. The inset shows the cross-section of the CDC region. 
}\label{fig:loop mirror design}
\end{figure*}

Achieving uniform mirror reflection that covers the entire laser gain spectrum is crucial for short pulse generation in integrated mode-locked lasers. In integrated photonic platforms, integrated loop mirrors can offer a broad reflection bandwidth. However, the reflectance of a loop mirror is determined by the coupling ratio of a directional coupler, which is typically wavelength dependent. To mitigate this wavelength dependence and achieve broader reflection bandwidth, in the loop mirror, we adopted a curved directional coupler (CDC) design as elaborated in Ref.\cite{mitarai2020design,morino2014reduction}. As shown in Fig. \ref{fig:loop mirror design}, our integrated loop mirror has a 50 $\mu$m-long CDC region with a bending radius of 600 $\mu$m.

\begin{figure*}[h!]
\centering
\includegraphics[width=1\linewidth]{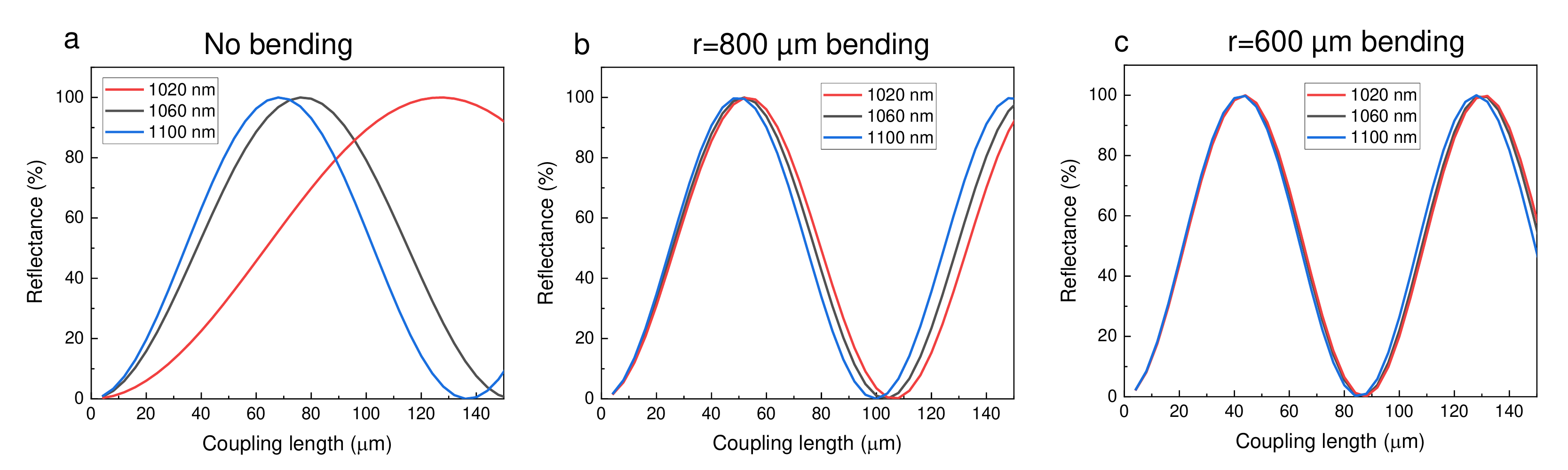}
\caption{Dependence of loop mirror reflectance on coupling length for (a) straight directional coupler, (b) curved directional coupler with 800 $\mu$m bending radius, and (c) curved directional coupler with 600 $\mu$m bending radius. 
}\label{fig:loop mirror performance}
\end{figure*}

Figure \ref{fig:loop mirror performance} compares the performance of loop mirrors with different directional coupler configurations. The operating wavelength range of interest is from 1020 to 1100 nm. As shown in Fig. \ref{fig:loop mirror performance}a, with a straight direction coupler, the reflectance of the loop mirror strongly depends on the operating wavelength. With a CDC of bending radius of 600 nm (Fig. \ref{fig:loop mirror performance}c), the loop mirror almost exhibits no wavelength dependence within the wavelength range of interest. The reflection reaches 100$\%$ when the coupling length is $\sim$45 $\mu$m.

\\

\\

\newpage
\section{Characterization of the electro-optic phase modulator}

The quantitatively analysis of our switching device including the nonlinear dynamics, the input/output coupling loss, the switching time and energy necessitates an accurate measurement of the input pulses. In Fig. \ref{fig:autocorrelation}a, we plot the measured spectrum of input pulses (red solid line). By comparing it with the spectra of 30-fs, 35-fs and 40-fs pulses centered at 2.09 $\mu$m, we found that the 35-fs pulse has the best agreement with our experimental spectrum. We also performed the interferometric autocorrelation measurement of the input pulses, as shown in Fig. \ref{fig:autocorrelation}b. The Gaussian fitting of the peaks of the autocorrelation has a FWHM of 65.2 fs, indicating that the actual pulse length is close to 46.2 fs. The slightly longer pulse length obtained from the autocorrelation measurement indicates that the input pulse is chirped, presumably due to the dispersive elements in our setup such as the pellicle, the long-pass filter and the neutral density (ND) filter. The relation between temporal profile of the pulse before ($a(t)$) and after ($a''(t)$) the dispersive element is given by\cite{weiner2011ultrafast}

\begin{align}
    a''(t)\approx\left(1+j\frac{\beta_2L}{2}\frac{d^2}{dt^2}\right)a(t)
\end{align}

\noindent where $L$ is length of the dispersive medium and $\beta_2$ is the group velocity dispersion (GVD) of the dispersive medium. Based on the results in Fig. S1a and b, we can estimate a total group dispersion delay (GDD) of $\beta_2L=\pm362$ fs$^2$. We determine the sign of GDD in section V.

\newpage
\section{Analysis of laser output pulse width}
 
In this section, we discuss the limiting factor of the output pulse width of our MLL by both the analytical solution of Haus's master equation and numerical simulation.

\subsection{Estimation of pulse width by Haus's master equation}

First, we analyze the fundamental limit of an actively MLL by considering the simplest case, which neglects the group delay dispersion (GDD) and nonlinear optical effects in the laser cavity, and the dynamic gain saturation. In this case, the output pulse width can be analytically solved from Haus's master equation. The Haus's master equation considering the  total pulse shaping due to gain, loss, and intra-cavity phase modulation can be written as:

\begin{align}
   T_R\frac{\partial A}{\partial T}=[\ g(T)+D_g \frac{\partial^2}{\partial T}-l-jM(1-\cos(2\pi f_\mathrm{M} t)] \ A
   \label{haus}
\end{align}

\noindent where $T_R$ is the round-trip time of the laser cavity, $g(T)$ is the time-dependent small-signal gain of the laser gain medium, $D_g$ is the group velocity dispersion (GVD) imposed by the gain medium, $M=\pi V_\mathrm{pp}/V_\pi$ is the modulation index of the phase modulator. We then simplify Eq. \ref{haus} by assuming a fixed small-signal gain value in the steady state.  The pulses, we expect, will have a width much shorter than the round-trip time $T_R$. They will be located in the minimum of the loss modulation where a parabolic approximation of the cosine function $\cos(2\pi f_\mathrm{M} t)\approx1-(2\pi f_\mathrm{M} t)^2/2$ and we obtain

\begin{align}
   T_R\frac{\partial A}{\partial T}=[\ g-l+D_g \frac{\partial^2}{\partial T}-j M_\mathrm{S} t^2] \ A
   \label{haus simp}
\end{align}

\noindent where $M_\mathrm{S}=M\omega_\mathrm{M}^2/2$ is the modulation strength and corresponds to the curvature of the phase modulation in the time domain at the minimum loss point. $D_g=g/( \pi \Delta f_\mathrm{g})^2$ is the gain dispersion\cite{keller2021ultrafast} and $\Delta f_\mathrm{g}$ is the 3 dB gain bandwidth. Eq. \ref{haus simp} is a linear partial differential equation, which can be solved by the separation of variables. The steady-state FWHM Gaussian pulse duration of an actively mode-locked laser with intracavity phase modulation is \cite{keller2021ultrafast}:

\begin{align}
   \tau_\mathrm{p}=0.53\sqrt[4]{\frac{g}{M}}\sqrt{\frac{1}{f_\mathrm{M} \Delta f_\mathrm{g}}}
\end{align}

\begin{figure*}[h!]
\centering
\includegraphics[width=0.8\linewidth]{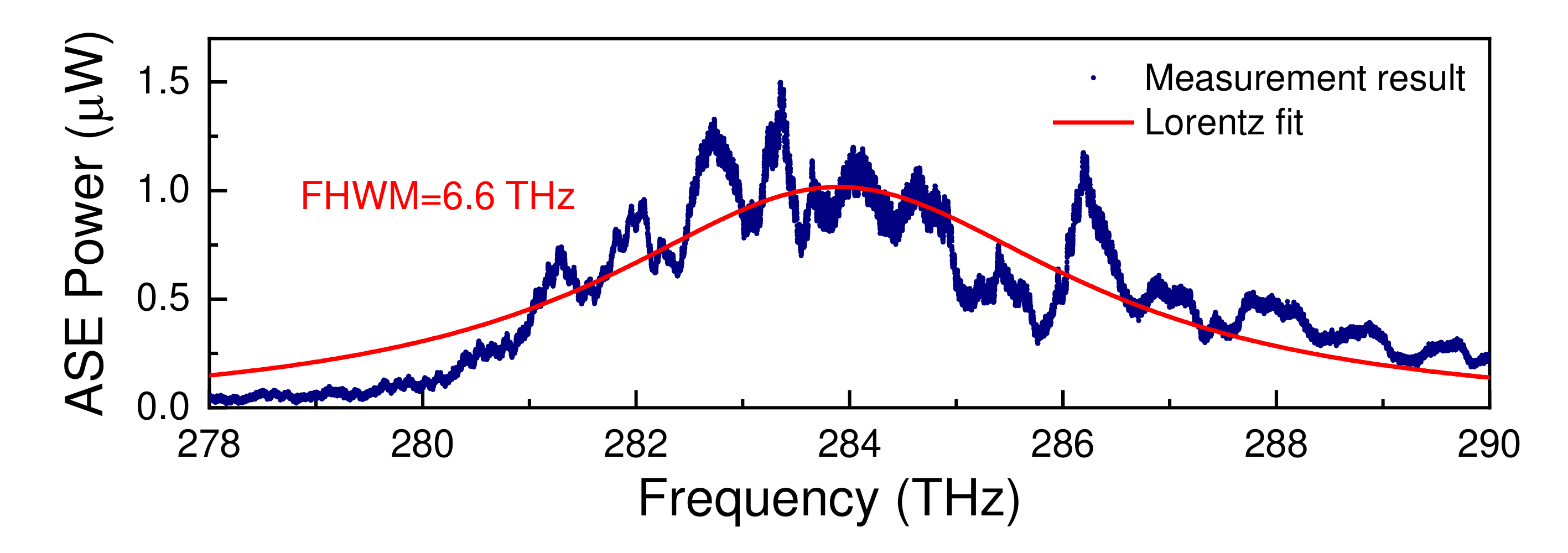}
\caption{Measured amplified spontaneous emission (ASE) spectrum of the semiconductor gain chip. The Lorentz fit of the ASE spectrum yields a gain bandwidth of 6.6 THz. 
}\label{fig:gain bandwidth}
\end{figure*}

As shown in Fig. \ref{fig:gain bandwidth}, by Lorentz fitting of the amplified spontaneous emission (ASE) spectrum of the semiconductor gain chip that we used in our experiments, we obtain a 3 dB gain bandwidth of 6.6 THz. Assuming the laser cavity has a round-trip $l_\mathrm{RT}$ loss of 70$\%$ (5.2 dB) due to chip coupling, output coupling, and propagation loss, in the steady state $2g=70\%$ so that $g$=0.35. Given the $V_\pi$ of our phase modulator is 10.67 V, with 280 mW RF driving power ($V_\mathrm{pp}=3.74$ V), the modulation index ($M$) is estimated to be 1.1. At $f_\mathrm{M}=10.17$ GHz, the estimated pulse width is:

\begin{align}
   \tau_\mathrm{p}\approx0.53\sqrt[4]{\frac{0.35}{1.1}}\sqrt{\frac{1}{10.17\times10^9\times6.6\times10^{12}}}=1.54\:\mathrm{ps}
\end{align}

\noindent which reflects the minimum pulse width we can in our actively mode-locked lasers. Small errors of $M$ and $g$ will not significantly change the resulting pulse width.

\subsection{Estimation of pulse width by numerical simulation}

We also compare the 

Now, we consider the more realistic case in actively mode-locked lasers, in which the GDD and the nonlinear optical effects in the laser cavity are taken into account.

\newpage




\bibliography{sup_refernces}